\documentclass[journal=jacsat,manuscript=article]{achemso}

\usepackage[version=3]{mhchem} 
\usepackage[english]{babel}
\usepackage{graphicx}
\usepackage{caption}
\usepackage{amsmath}
\usepackage{subcaption}
\usepackage{bm}
\pdfoutput=1 

\author{Andrew Abi Mansour}
\author{Peter J. Ortoleva}
\email{ortoleva@indiana.edu}
\affiliation{Department of Chemistry, Indiana University, Bloomington}

\title[\texttt{achemso} demonstration]
{A Multiscale Factorization Method for Simulating
Mesoscopic Systems with Atomic Precision}

\begin{document}

\begin{abstract}
Mesoscopic $N-$atom systems derive their structural and dynamical properties
from processes coupled across multiple scales in space and time. An
efficient method for understanding these systems in the friction dominated regime from the
underlying $N-$atom formulation is presented. The method integrates notions
of multiscale analysis, Trotter factorization, and a hypothesis that the 
momenta conjugate to coarse-grained variables can be treated as a stationary 
random process. The method is demonstrated for Lactoferrin, Nudaurelia 
Capensis Omega Virus, and Cowpea Chlorotic Mottle Virus to assess
its accuracy and scaling with system size.
\end{abstract}

\section{Introduction}

The objective of the present study is to simulate the behavior of mesoscopic
systems based on an all-atom formulation at which the basic Physics is presumed
known. Traditional molecular dynamics (MD) is ideal for such an approach if
the number of atoms and the timescales of interest are limited \cite{NAMD,GROMACS}. 
However, ribosomes, viruses, mitochondria, and nanocapsules for the
delivery of therapeutic agents are but a few examples of mesoscopic systems
that can provide a challenge for conventional MD. In this paper, we develop a 
Physics-based algorithm that accounts for interactions at the atomic scale 
and yet makes accurate and rapid simulations for supramillion atom systems over 
long timescales possible.

Typical coarse-graining (CG) methods include deductive multiscale 
analysis (DMA) \cite{Thermal,DMA}, inverse Monte Carlo \cite{IMC}, 
Boltzmann inversion \cite{IBM}, elastic network models \cite{ENM1,ENM2}, 
or other bead-based models \cite{CG1,CG2,Martini}. DMA methods derived 
from the $N-$atom Liouville equation (LE) show
great promise in achieving accurate and efficient all-atom simulation \cite
{OPs, Peter2005, Smoluchowski, Langevin}. The main theme of that work was to
construct and exploit the multiscale structure of the $N-$atom probability
density $\rho (\Gamma ,t)$ for the positions and momenta of the $N$ atoms
(denoted $\Gamma $, collectively) as it evolves over time $t$. Most of the analysis
focused on friction dominated, non-inertial regime, which is considered here as well.
However, in these methods ensembles of all-atom configurations were required for evolving 
the CG variables. The 
approach introduced here avoids the need to construct these ensembles by
coevolving the all-atom and CG states in a consistent way, and in the spirit of DMA-based methods,
it does not make any conjectures on the form of the CG dynamical equations and the associated
uncertainty in the form of the equations. A main theme of the present approach is the importance of coevolving the CG and microscopic states. This feature distinguishes our method from others which, for example, require the construction of a potential mean force \cite{Tuckerman2008, Voth2008} using ensembles of micostates; a challenge for such methods is that the relevant ensembles are not known a priori since they are controlled by the CG state whose evolution is unknown, and is in fact the objective of a dynamics simulation. As a result, the present method does not require least squares or other types of fitting. Other multiscale methods, built on the projection operator formalism \cite{Oppenheim1996,Oppenheim1997,Oppenheim1998}, require the construction of memory kernels. This is typically achieved via a perturbation approach to overcome the complexity of the appearance of the projection operator in the memory kernels. Construction of such kernels is not required in our method. 

A first step in the present approach is
the introduction of a set of CG variables $\Phi$ related to $\Gamma$ via 
$\Phi =\overset{\sim}{\Phi } (\Gamma )$ for specified function 
$\overset{\sim}{\Phi } (\Gamma )$. When
this dependence is well chosen,
the CG variables evolve much more slowly than the fluctuations of small
subsets of atoms. With these CG variables, the $N-$atom
LE was solved perturbatively in terms
of $\varepsilon$\cite{OPs, Peter2005}, the ratio of the characteristic time of the fluctuations
of small clusters of atoms, to the characteristic time of CG variable
evolution. This is achieved starting with the ansatz that $\rho$ depends on 
$\Gamma$ both directly and, via $\overset{\sim}{\Phi}$, indirectly. The
theory proceeds by constructing $\rho \left( \Gamma ,\Phi ;t\right)$
perturbatively in $\varepsilon$, i.e., by working in the space of functions
of $6N+N_{CG}$ variables (where $N_{CG}$ is the number of variables in the
set $\Phi$). To advance the multiscale approach, we here introduce Trotter factorization 
\cite{Trotter,Algebra,Tuckerman} into the analysis. Through
Trotter factorization, the long-time evolution of the system separates into
alternating phases of all-atom simulations and CG variable updating.
Efficiency of the method follows from a hypothesis that the momenta
conjugate to the CG variables can be represented as a stationary random
process. The net result is a computational algorithm with some of the 
character of our earlier MD/OPX method \cite{Long1,Long2} but with 
greater control on accuracy, higher efficiency, and more rigorous theoretical 
basis. Here we develop the algorithm and discuss its implementation as a 
computational platform, discuss selected results, 
and make concluding remarks.

\section{Theory and Implementation}

\subsection{\protect\bigskip Unfolded Dynamical Formulation}
The Newtonian description of an $N-$atom system is provided by the $6N$ atomic positions and momenta, denoted $\Gamma$
collectively. The phenomena of interest involve overall transformations of an $N-$atom system.
While $\Gamma$ contains all the information needed to solve the problem in principle, here it is found convenient to also introduce a set of CG variables $\Phi$, that are used to track the large spatial scale, long time degrees of freedom. For example, $\Phi$ could describe the overall position, size,
shape, and orientation of a nanoparticle. By
construction, a change in $\Phi$ involves the coherent deformation of the $
N-$atom system, which implies that the rate of change in $\Phi$ is expected
to be slow \cite{OPs,Joshi2012}.  This slowness implies the separation of timescales that provides a highly efficient and accurate algorithm for simulating $N-$atom systems.  

With this unfolded description $(\Gamma, \Phi)$, the Newtonian dynamics takes the form
\begin{align}
\frac{d \Gamma }{d t}&=\mathcal{L} \Gamma, \label{dynamics_gamma} \\
\frac{d \Phi }{d t}&=\mathcal{L} \overset{\sim}{\Phi}(\Gamma), \label{dynamics_phi}
\end{align}
for unfolded Liouvillian $\mathcal{L} = \mathcal{L}_{micro} + \mathcal{L}_{meso}$, such that
\begin{align}
\mathcal{L}_{micro} &= \sum_{l=1}^N \frac{\mathbf{p}_l}{m_l} \cdot \left( \frac{\partial}{\partial\mathbf{r}_l}\right)_{\Phi} + \mathbf{f}_l \cdot \left( \frac{\partial}{\partial \mathbf{p}_l} \right)_{\Phi}, \\
\mathcal{L}_{meso} & = \sum_{k=1}^{N_{CG}} \Pi_k \cdot \left( \frac{\partial}{\partial \Phi_k} \right)_{\Gamma}. 
\end{align}
Here $\Pi_k$ is the CG velocity associated with the $k^{th}$ CG variable. Eqs. (\ref{dynamics_gamma}-\ref{dynamics_phi}) have the formal solution
\begin{equation}
(\Gamma(t), \Phi(t)) = S(t) (\Gamma_o, \Phi_o),
\end{equation}%
for initial data indicated by subscript $o$, and evolution operator $S(t)=e^{\mathcal{L} t}$.
 
\subsection{Trotter Factorization}

By taking the unfolded Liouvillian, the time operator now takes the form 
\begin{equation}
S(t)=e^{\left( \mathcal{L} _{{micro}} +\mathcal{L} _{{meso}} \right)t}.
\end{equation}%
Since $\mathcal{L} _{{micro}}$ and $\mathcal{L} _{{meso}}$ do not
commute, $S(t)$ cannot be factorized into a product of exponential
functions. However, Trotter's theorem \cite{Trotter} (also known as the Lie
product formula \cite{Algebra}) can be used to factorize the evolution operator 
as follows: 
\begin{equation}
S(t)=\lim_{M\rightarrow \infty }\left[ e^{\mathcal{L} _{{micro}}t/2M}e^{%
\mathcal{L} _{{meso}}t/M} e^{\mathcal{L} _{{micro}}t/2M}\right]
^{M}+O\left( \left(\frac{t}{M}\right)^3 \right) .
\end{equation}%
By setting $t/M$ to be equal to the discrete time step $\Delta$, the step-wise
operator becomes 
\begin{equation}
S(\Delta )=\lim_{\Delta \rightarrow 0} e^{\mathcal{L} _{{micro}%
}\Delta/2} e^{ \mathcal{L} _{{meso}}\Delta }e^{\mathcal{L} _{{micro}%
}\Delta/2} + O(\Delta^3).
\end{equation}%
Let the step-wise operators $S_{{micro}}$ and $S_{{meso}}$
correspond to $\mathcal{L}_{{micro}}$ and $\mathcal{L}_{{meso}}$,
respectively. Then $S(n\Delta)$ takes the form%
\begin{equation}
S(n\Delta )= \prod_{i=1}^n S_{{micro}}\left(\frac{\Delta}{2}\right) S_{%
{meso}}(\Delta) S_{{micro}}\left(\frac{\Delta}{2}\right).
\label{Full_S}
\end{equation}
By replacing $S_{{micro}}(\Delta/2)$ by $S_{{micro}}(\Delta) S_{%
{micro}}(-\Delta/2)$ to the right hand side, Eq. (\ref{Full_S}) becomes 
\begin{equation}
S(n\Delta )= S_{micro}\left(\frac{\Delta}{2}\right) \left[ \prod_{i=1}^n S_{%
meso}(\Delta) S_{micro}\left(\Delta\right) \right] S_{micro} 
\left(\frac{-\Delta}{2}\right).
\end{equation}
Since we are interested in the long-time evolution of a mesoscopic system,
we can neglect the far left and right end terms, $S_{micro}\left({%
\Delta}/2\right)$ and $S_{micro}\left({-\Delta}/2\right)$,
respectively, to a good approximation. Therefore, we can define the
step-wise time operator as 
\begin{equation}
S\left( \Delta \right) =S_{meso}\left( \Delta \right) S_{micro}
\left( \Delta \right).
\end{equation}%
In the next section, we show how this factorization implies a computational
algorithm for solving the dynamical equations for $\Gamma$ and $\Phi$.

\subsection{Implementation}
A key to the efficiency of the mutiscale Trotter factorization (MTF) method
is the postulate that the momenta conjugate to the CG variables can be
represented by a stationary random process over a period of time much
shorter than the time scale characteristic of CG evolution. Thus, in a time
period significantly shorter than the increment $\Delta$ of the step-wise
evolution, the system visits a representative ensemble of configurations
consistent with the slowly evolving CG state. This enables one to use an MD
simulation for the microscopic phase of the step-wise evolution that is much
shorter than $\Delta$ to integrate the CG state to the next CG time step.
For each of a set of time intervals much less than $\Delta$, the friction
dominated system experiences the same ensemble of conjugate momentum 
fluctuations. Thus, if $\delta$ is the time for which the conjugate momentum 
undergoes a representative sample of values (i.e., is described by the stationarity
hypothesis), then the computational advantage over conventional MD is
expected to be $\Delta/\delta$.

The two phase updating for each time-step $\Delta$ was achieved as follows.
For the $S_{micro}(\Delta)$ phase, conventional MD was used. This yields 
a time-series for $\Gamma$ and hence ${\Pi}$. For all systems simulated
here, $\Pi$ was found to be
a stationary random process (see Figure \ref{fig:stationarity}). Therefore,
MD need only be carried out for a fraction of $\Delta$, denoted
$\delta$. This and the slowness of the CG variables are the source of computational 
efficiency of our algorithm. For the $S_{meso}$ phase updating in the friction dominated regime, the ${\Pi}$ time 
series constructed in the micro phase is used to advance $\Phi$ in time as follows
\begin{equation}
\Phi(t+\Delta) = \Phi(t) + \int_t^{t+\Delta} dt' \Pi(t'). \label{integrated_phi}
\end{equation}
Due to stationarity, the integral on the right hand side reduces to
$\Delta/\delta \int_t^{t+\delta} dt' \Pi(t')$ (see Figure \ref{fig:stationarity}). 
The expression for $\Pi$
depends on the choice of CG variables. In this work, we used the space-warping method \cite{Khuloud2002,Pankavich2008}
that maps a set of atomic coordinates to a set of CG variables that capture the coherent deformation 
of a molecular system in space. In the space-warping method, the  mathematical relation between the 
CG variables and the atomic coordinates is
\begin{equation}
\mathbf{r}_i = \sum_{\underline{k}} \mathbf{U}_{\underline{k}i} \bm{\Phi}_{\underline{k}} + \mathbf{\sigma}_i. \label{Map}
\end{equation}
Here $\underline{k}$ is a triplet of indices, $i$ is the atomic index, $\mathbf{r}_i$ is the cartesian position vector for atom $i$, and $\bm{\Phi}_{\underline{k}}$ is a cartesian vector for CG variable $\underline{k}$. The basis functions $\mathbf{U}_{\underline{k}}$ are constructed in two stages. In the first stage, they are computed from a product of three Legendre polynomials of order $k_1$, $k_2$, and $k_3$ for the $x$, $y$, and $z$ dependence. In the second stage, the basis functions are mass-weighted orthogonalized via QR decomposition \cite{OPs,Joshi2012}. For instance, the zeroth order polynomial is $\mathbf{U}_{000}$, the first order polynomial forms a set of three basis functions: $\mathbf{U}_{001}, \mathbf{U}_{010}, \mathbf{U}_{100}$, and so on. Furthermore, the basis functions depend on
a reference configuration $\mathbf{r}^0$ which is updated periodically (once every $10$ CG time steps) to control accuracy. 
The vector $\mathbf{\sigma}_i$ represents the atomic-scale corrections to the coherent deformations generated by $\bm{\Phi}_{\underline{k}}$. Introducing CG variables this way facilitates the construction of microstates consistent with the CG state \cite{Pankavich2008}. This is achieved by minimizing $\sum_{i=1}^N m_i \mathbf{\sigma}_i^2$ with respect
to $\bm{\Phi}_{\underline{k}}$. The result is that the CG variables are generalized centers of mass, specifically
\begin{equation}
\bm{\Phi}_{\underline{k}} = \frac{\sum_{i=1}^N m_i \mathbf{U}_{\underline{k}i} \mathbf{r}_i}{\sum_{i=1}^N m_i \mathbf{U}_{\underline{k}i}^2}, \label{CG_Eq}
\end{equation}
with $m_i$ being the mass of atom $i$. For the lowest order CG variable, $\mathbf{U}_{000}=1$, which implies $\bm{\Phi}_{000}$ is the center of mass. As the order of the polynomial increases, the CG variables capture more information from the atomic scale, but they vary less slowly with time. Therefore, the space warping CG variables are classified into low order and high order variables. The former characterize the larger scale disturbances, while the latter capture short-scale ones \cite{OPs,Joshi2012}. Eq. (\ref{CG_Eq}) implies that
$\bm{\Pi}_{\underline{k}}= \sum_{i=1}^{N} \mathbf{U}_{\underline{k}i} \mathbf{p}_i / \sum_{i=1}^N m_i \mathbf{U}_{\underline{k}i}^2$, where $\mathbf{p}_i$ is a vector of momenta
for the $i^{th}$ atom. With $\bm{\Phi}(t+\Delta)$ computed via Eq. (\ref{integrated_phi}), the two-phase $\Delta$ update
is completed, and this cycle is repeated for a finite number of discrete time steps. Details on
the necessary energy minimization and equilibriation needed
for every CG step was covered in earlier work \cite{OPs,Long1,Long2}.
This two-phase coevolution algorithm was implemented using NAMD 
\cite{NAMD} for the $S_{micro}$ phase within the framework of the DMS software package \cite{OPs,History,Thermal}. 
Numerical computations were performed with 
the aid of LOOS \cite{LOOS}, a lightweight object-oriented structure library.

\section{Results and discussion}

\begin{figure}[tbp]
\centering
\begin{subfigure}[b]{0.7\textwidth}
                \centering
                \includegraphics[width=\textwidth]{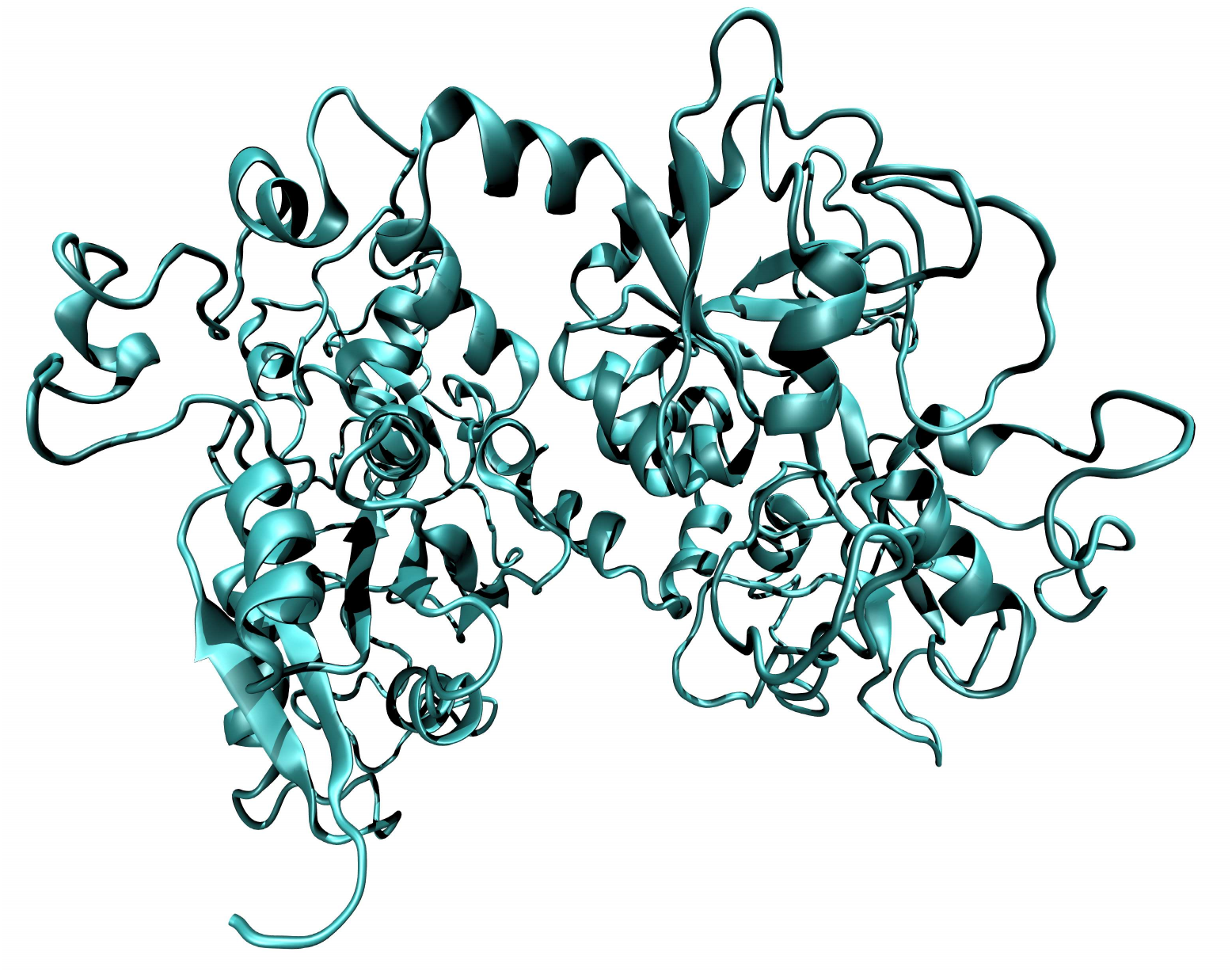}
                \caption{LFG in its open state at $t=0$ ns.}
                \label{fig:1LFG-open}
\end{subfigure}
~
\begin{subfigure}[b]{0.7\textwidth}
                \centering
                \includegraphics[width=\textwidth]{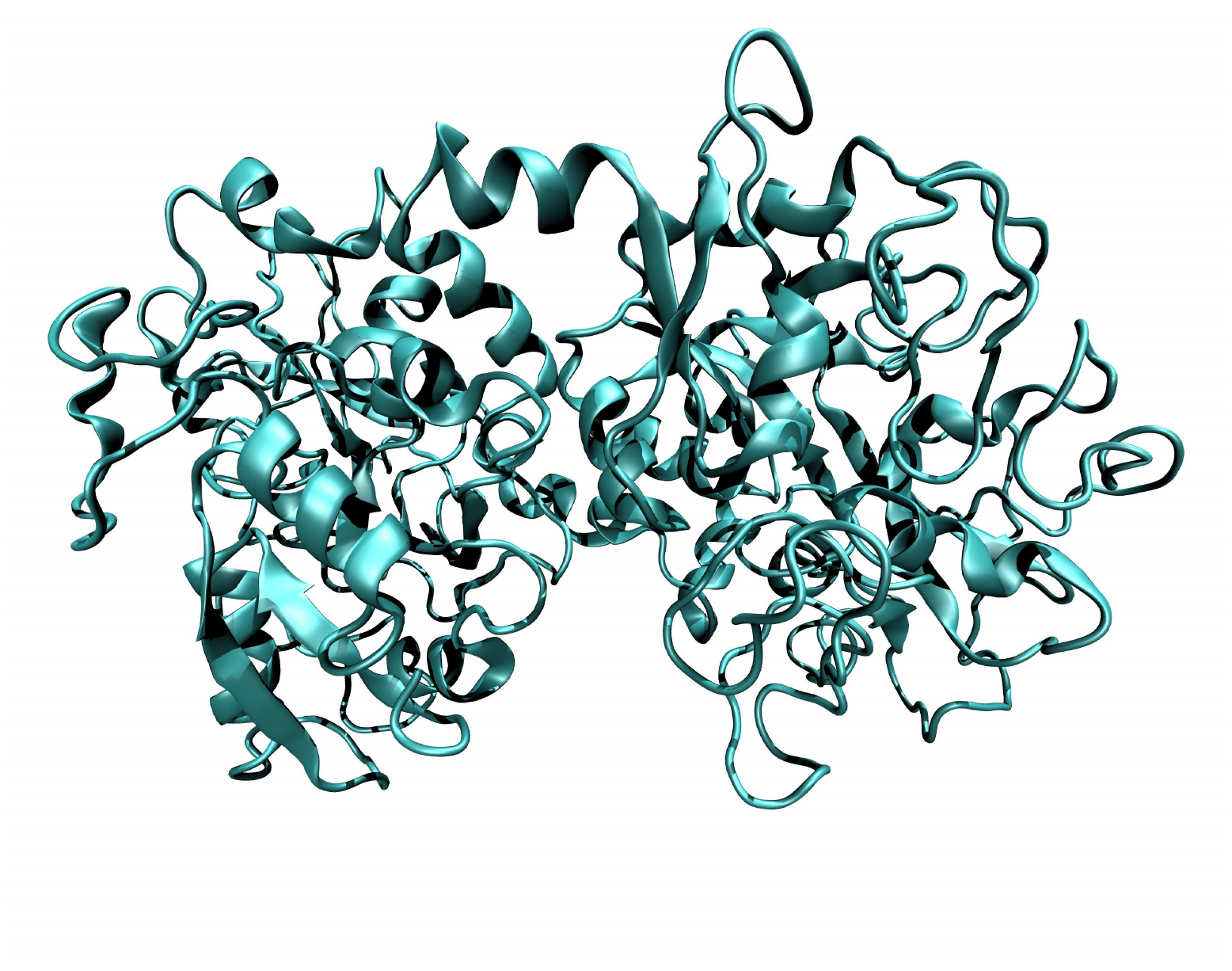}
                \caption{LFG in its closed state at $t=19.6$ ns.}
                \label{fig:1LFG-closed}
                \end{subfigure}
\caption{Snapshots of Lactoferrin protein in its open (a) and closed
(b) states.}
\label{fig:lfg}
\end{figure}

\begin{figure}[tbp]
\centering
\includegraphics[width=\textwidth]{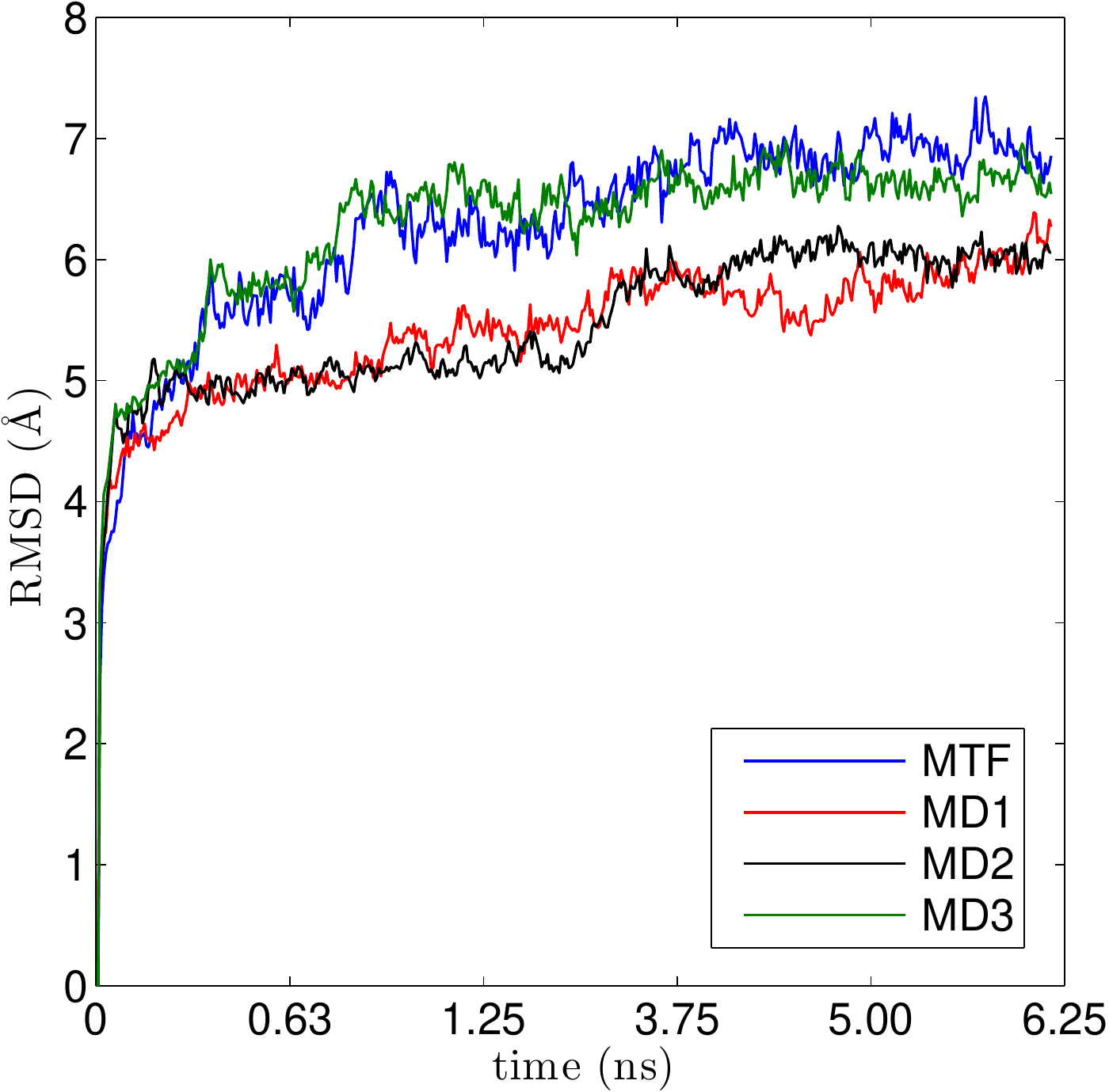}
\caption{RMSD variation as a function of time for a series of three MD and one MTF runs.}
\label{fig:rmsd-1LFG}
\end{figure}

\begin{figure}[tbp]
\includegraphics[width=\textwidth]{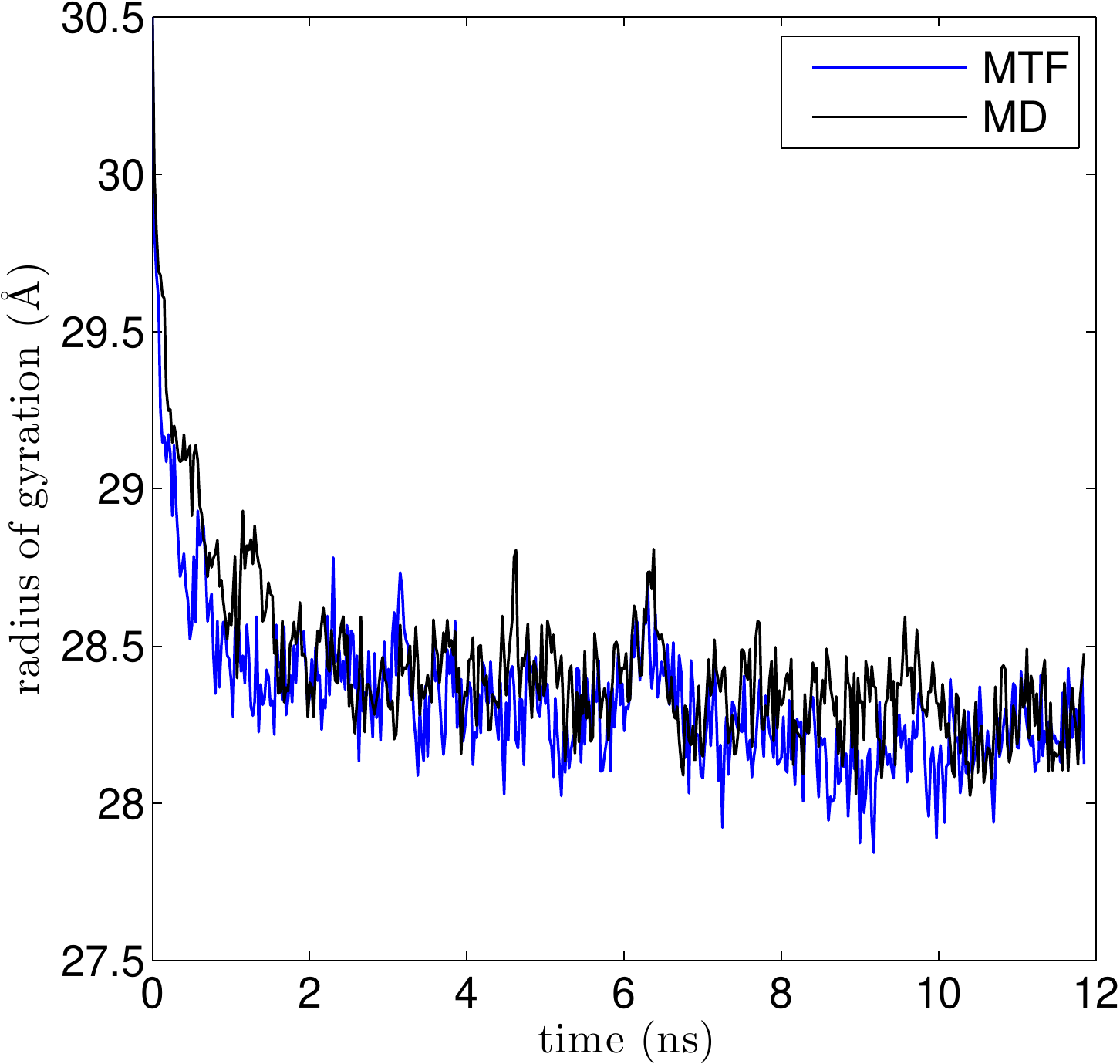}
\centering
\caption{The radius of gyration decreases in time as Lactoferrin shrinks.}
\label{fig:rog-1LFG}
\end{figure}

All simulations were done in vacuum under NVT conditions to assess the scalability and accuracy of the algorithm.
The first system used for validation and benchmarking is lactoferrin. This
iron binding protein is composed of a distal and two proximal lobes (shown in Figure \ref{fig:1LFG-open}). 
Two free energy minimizing conformations have been
demonstrated experimentally: diferric with closed proximal lobes (PDB code
1LFG), and apo with open ones \cite{1LFG1} (PDB code 1LFH). Here, we start 
with an open lactoferrin structure and simulate its closing in vacuum (see Figure \ref{fig:lfg}). 
The RMSD for Lactoferrin is plotted as a function of
time in Figure \ref{fig:rmsd-1LFG}; it shows that the protein reaches
equilibrium in about $5$ ns. This transition leads to a decrease
in the radius of gyration of the protein by approximately $0.2$ nm as shown in Figure \ref{fig:rog-1LFG}.

\begin{figure}[tbp]
\centering
\begin{subfigure}[b]{0.45\textwidth}
                \centering
                \includegraphics[width=\textwidth]{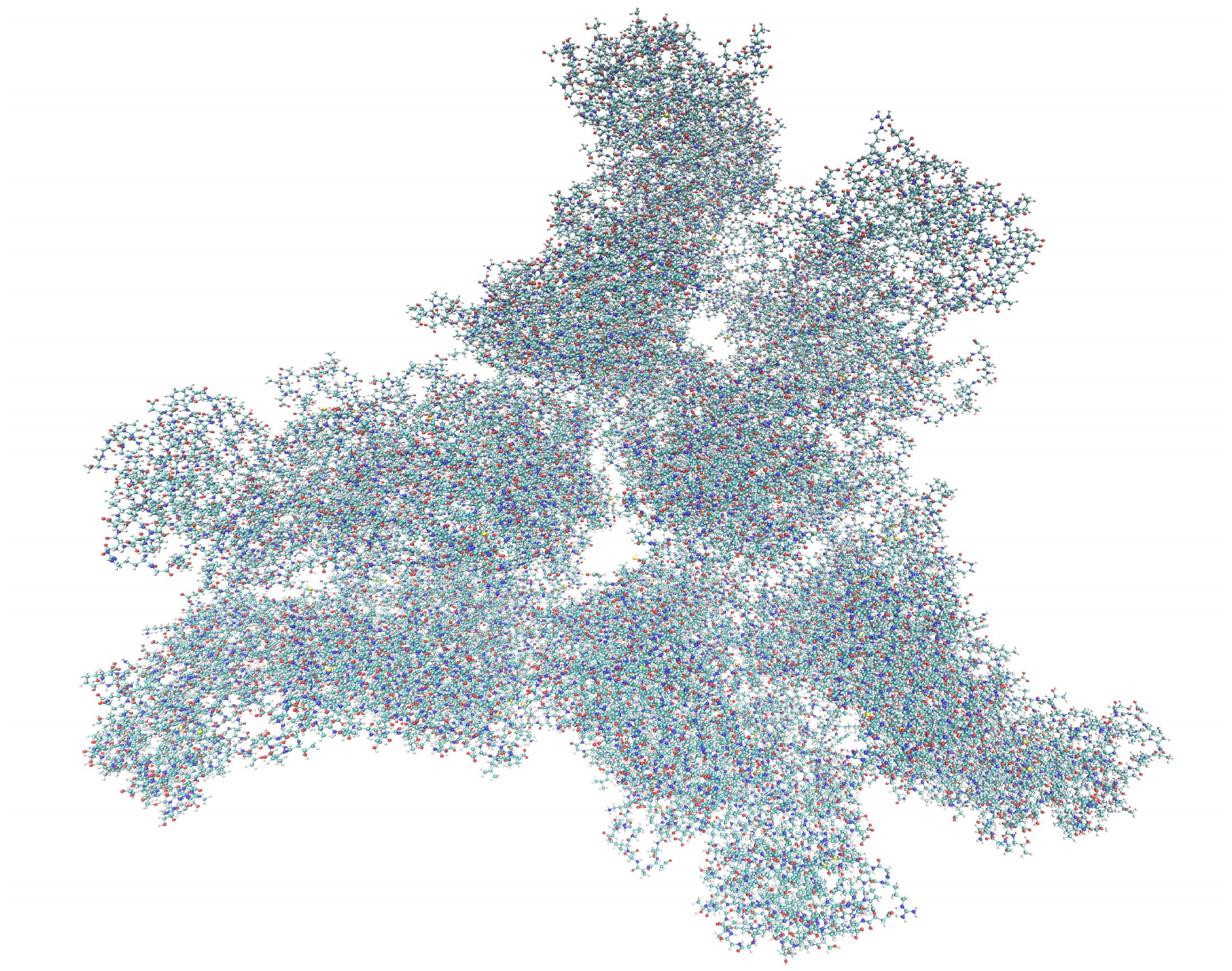}
                \caption{N$\omega$v in its initial state at $t=0$ ns.}
                \label{fig:1nwv-open}
        \end{subfigure}
~
\begin{subfigure}[b]{0.45\textwidth}
                \centering
                \includegraphics[width=\textwidth]{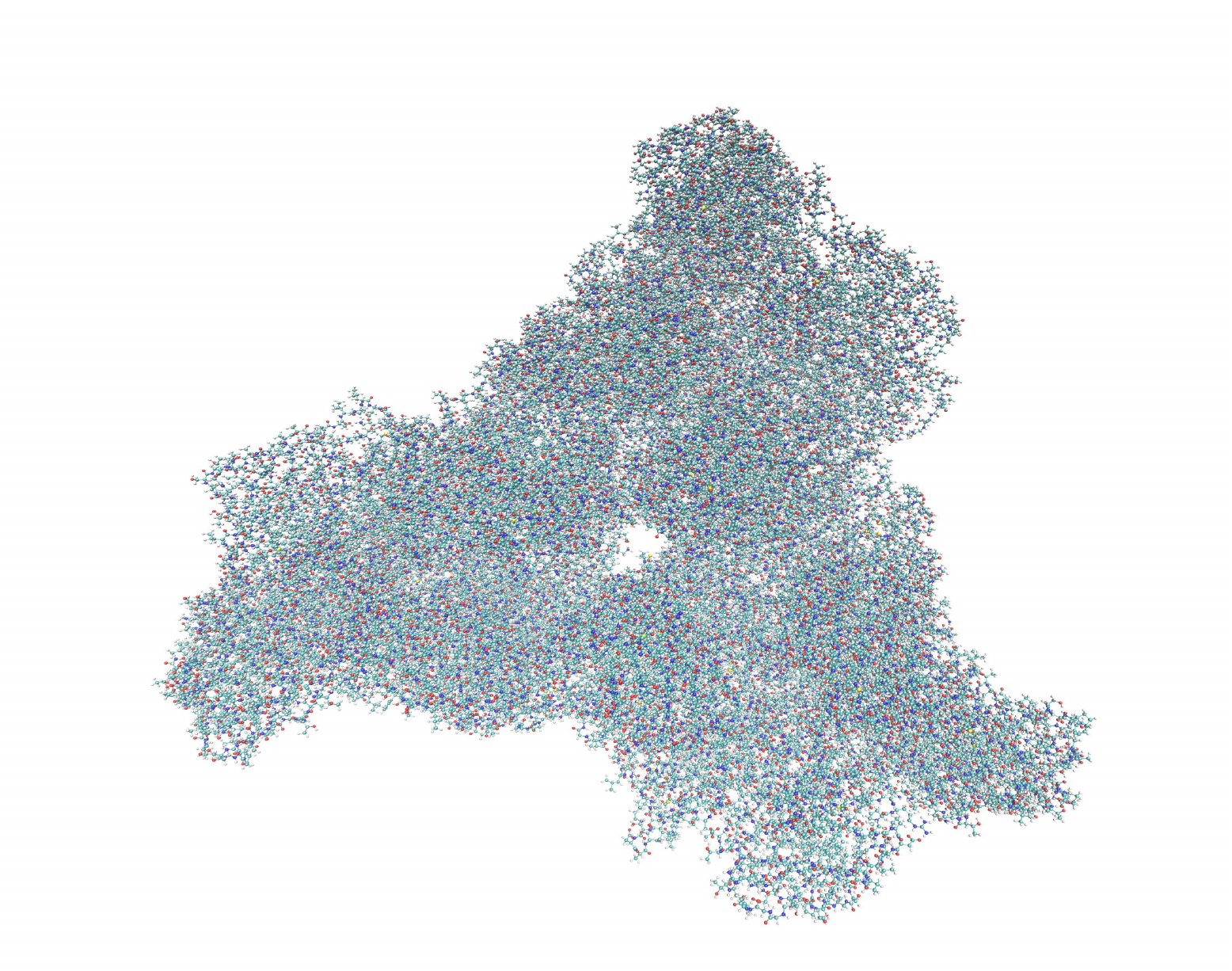}
                \caption{N$\omega$v after shrinking at $t=3.0$ ns.}
                \label{fig:1nwv-closed}
        \end{subfigure}
\caption{Snapshots of N$\omega$v triangular structure before (a) and after (b) contraction due to
strong protein-protein interactions.}
\label{fig:ohf}
\end{figure}

\begin{figure}[tbp]
\centering
\includegraphics[width=\textwidth]{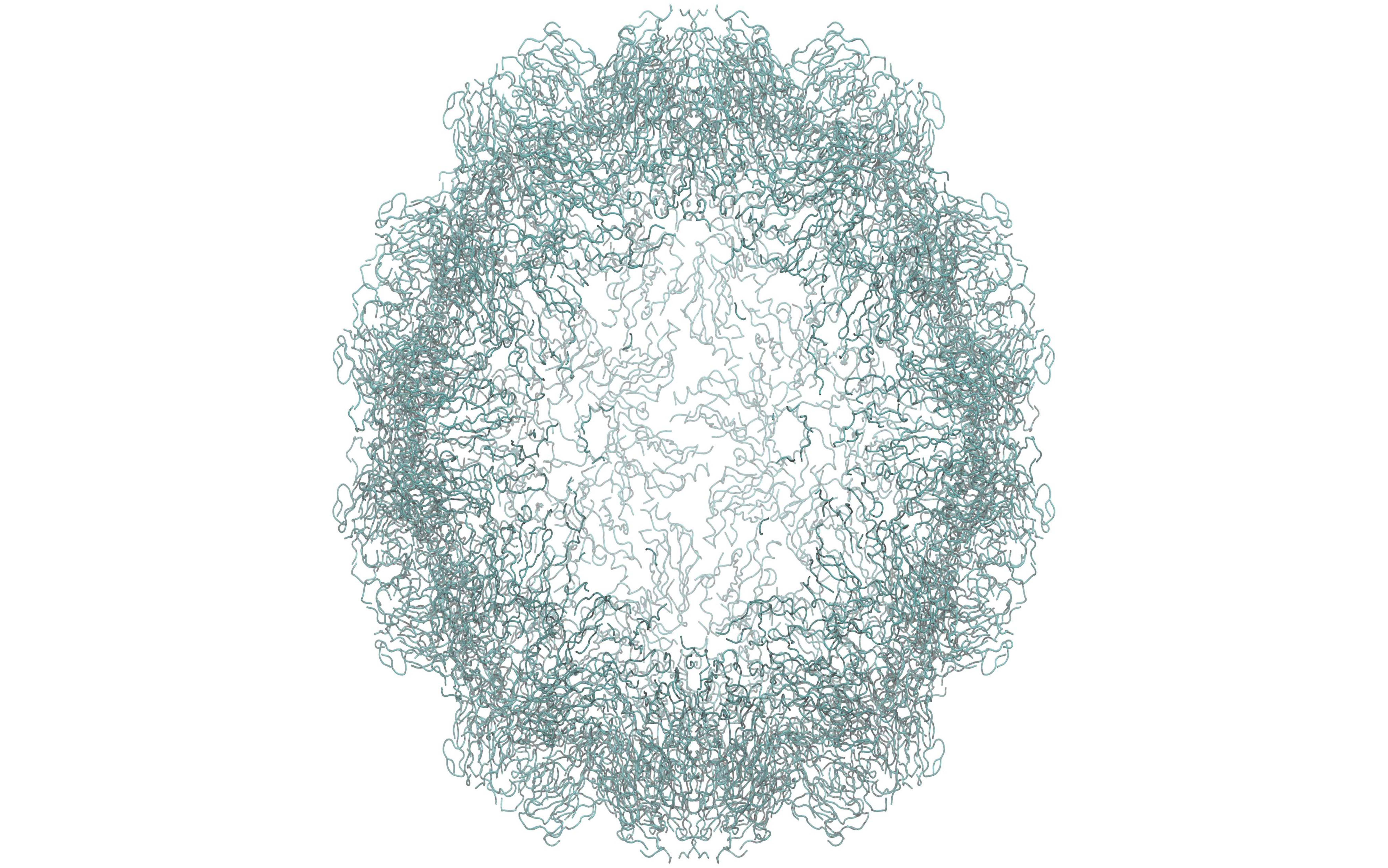}
\caption{The full all-atom CCMV native capsid.}
\label{fig:cwp}
\end{figure}

\begin{figure}[tbp]
\centering
\includegraphics[width=\textwidth]{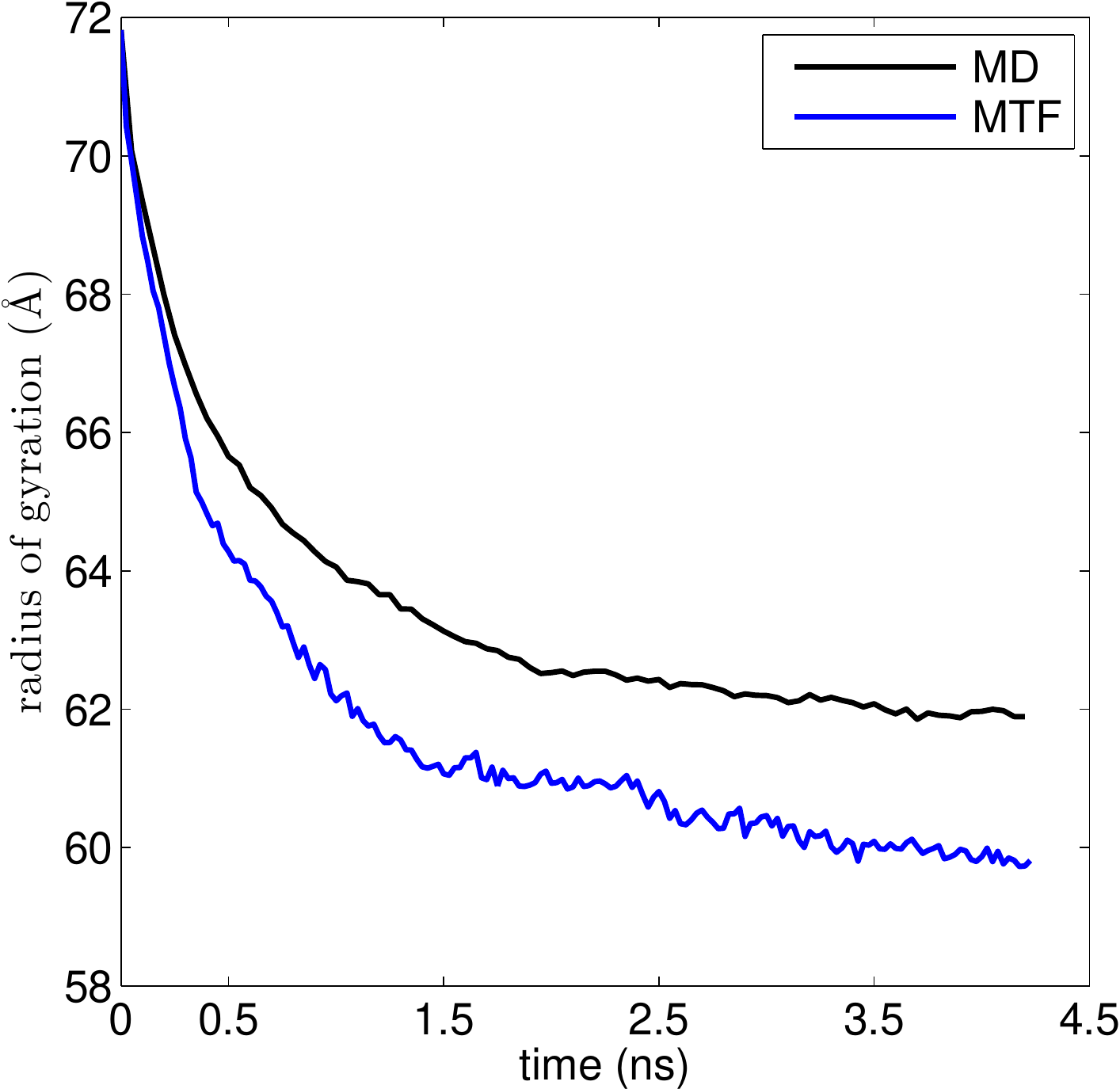}
\caption{Temporal evolution of the radius of gyration of N$\omega$v computed using MD and MTF.}
\label{fig:rog-ohf}
\end{figure}

\begin{figure}[tbp]
\centering
\includegraphics[width=\textwidth]{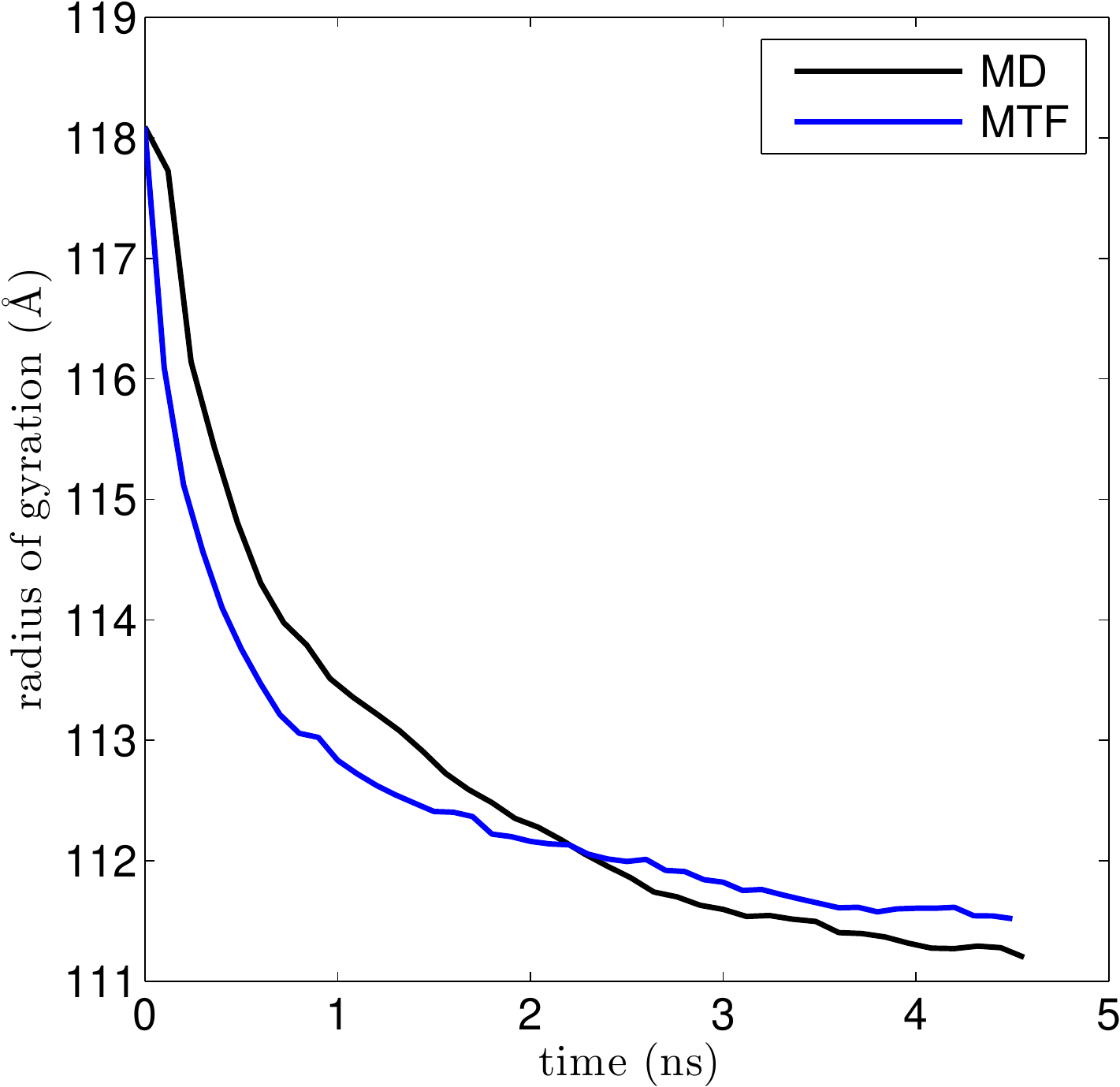}
\caption{Temporal evolution of the radius of gyration of CCMV computed using MD and MTF.}
\label{fig:rog-cwp}
\end{figure}

The second used is a triangular structure of the
Nudaurelia Capensis Omega Virus (N$\omega$V) capsid protein \cite{Nwv} (PDB code 1OHF) containing
three protomers (see Figure \ref{fig:ohf}). Starting from a deprotonated state (at low pH), the system
was equilibriated using an implicit solvent. The third system used is Cowpea Chlorotic Mottle virus (CCMV) 
full native capsid \cite{Long1} (PDB code 1CWP, Figure \ref{fig:cwp}). Both systems are characterized by 
strong protein-protein interactions. As a result, they shrink in vacuum
after a short period of equilibriation. The computed radius of gyration of both systems is shown
in Figure \ref{fig:rog-ohf} and Figure \ref{fig:rog-cwp}.

\begin{figure}[tbp]
\centering
\begin{subfigure}[b]{0.55\textwidth}
                \centering
                \includegraphics[width=\textwidth]{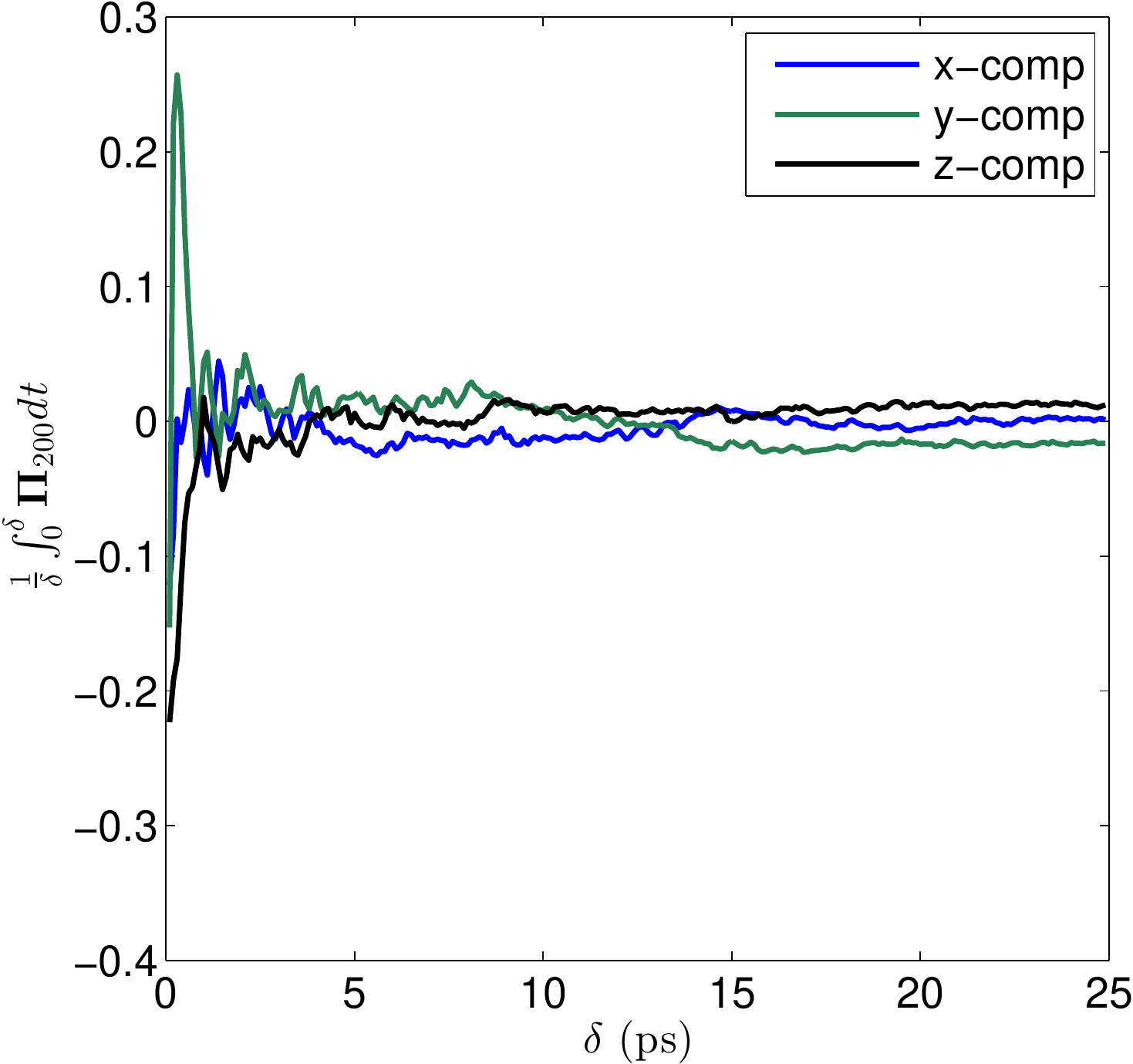}
                \caption{A plot of the time integral of $\Pi$ for a high order CG $\Phi_{200}$.}
\end{subfigure}
~

\begin{subfigure}[b]{0.55\textwidth}
        \centering
        \includegraphics[width=\textwidth]{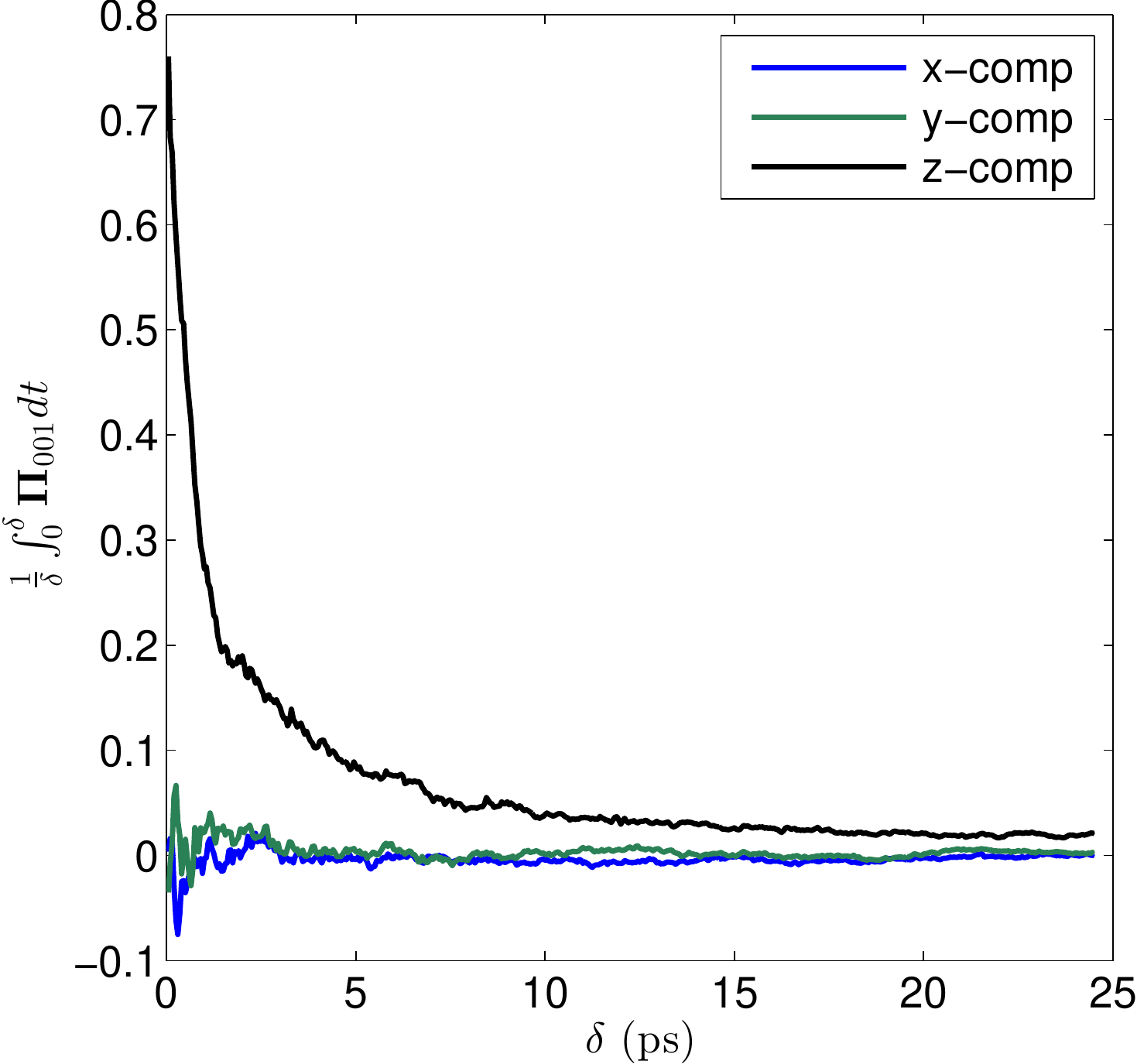}
        \caption{A plot of the time integral of $\Pi$ for a low order CG $\Phi_{001}$.}
\end{subfigure}
\caption{Evidence for the validity of the stationarity hypothesis shown via the convergence of $\frac{1}{\delta} \int_0^{\delta} \Pi(t) dt$ as a function of $\delta$ for CG variables selected from among those used in simulating the contraction 
of N$\omega$v. Initially the integral experiences large fluctuations because with small $\delta$, only a relatively few configurations are included in the time average constituting the integral, but as $\delta$ increases, the statistics improves, and the integral becomes increasingly flat.}
\label{fig:stationarity}
\end{figure}

Based on the convergence of the time integral of $\Pi$ (see Figure \ref{fig:stationarity}), the $S_{micro}$ phase was chosen to
consist of $10 × \times 10^4$ MD steps for LFG, N$\omega$v, and CCMV, where each MD step
is equal to $1$ fs. The CG timestep, $\Delta$, on the other hand, was taken to be $12.5$ ps for LFG, $25$ ps for
N$\omega$v, and $50$ ps for CCMV.

\begin{figure}[tbp]
\centering
\includegraphics[width=\textwidth]{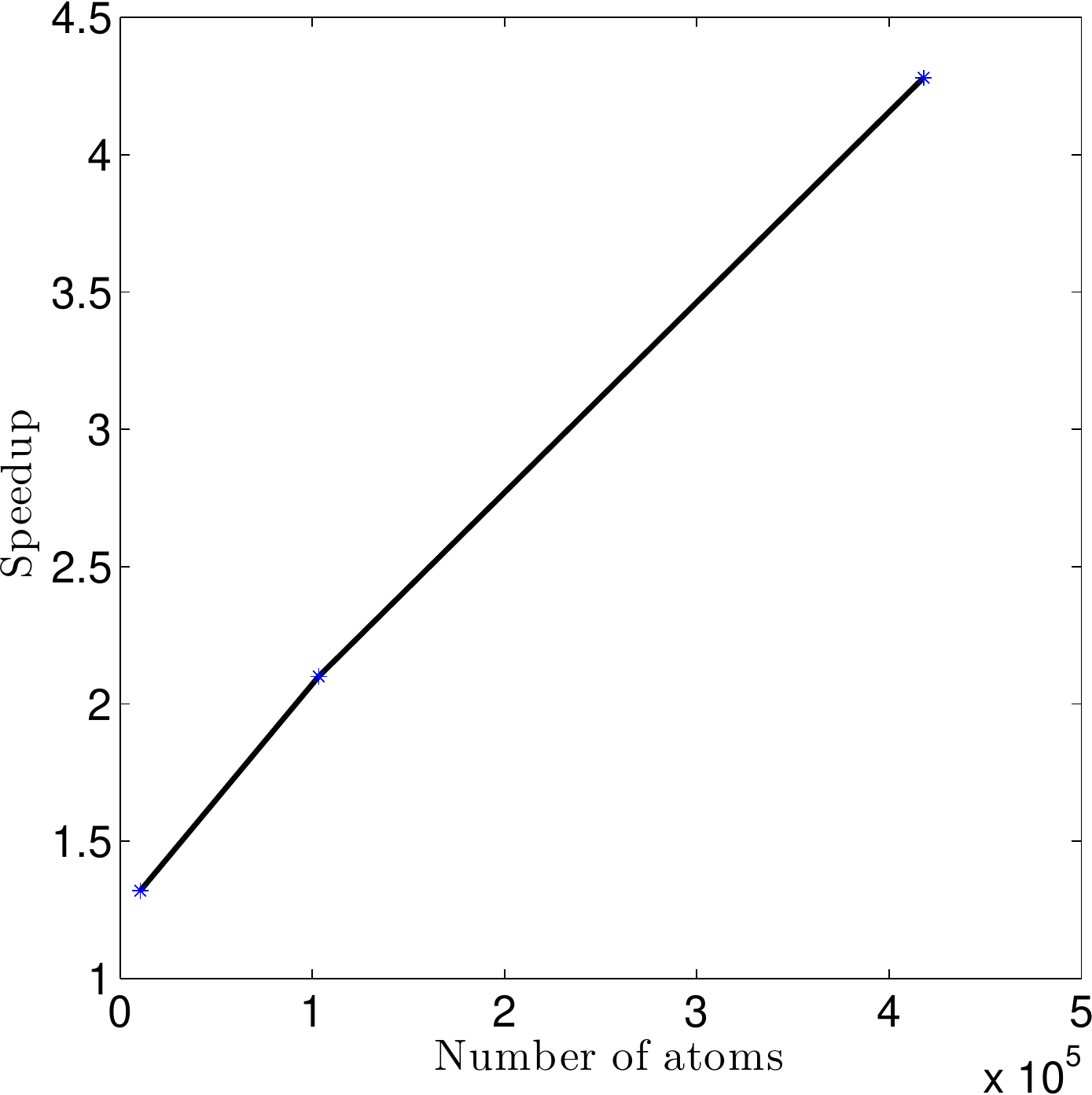}
\caption{A plot of the speedup as a function of the system size shows the scalability of the MTF algorithm.}
\label{fig:speedup}
\end{figure}

\begin{table}
\begin{center}
\begin{tabular}{|l|l|l|l|l|l|}
\hline
System & Size & Time & Speed-up \\ \hline
LFG & 10,560 & 12.5 ns & 1.32 \\ \hline
N$\omega$V & 103,317 & 4.30 ns & 2.10 \\ \hline
CWP & 417,966 & 4.67 ns & 4.28 \\ \hline
\end{tabular}
\caption{Speedup as a function of system size (number of atoms) for simulations run on 1x12, 4x12, and 8x12 
cores for LFG, N$\omega$v, and CCMV, respectively.}
\label{tab:speedup}
\end{center}
\end{table}
Comparison between MD and MTF results are shown in \ref{tab:speedup}. The dependence 
of speedup on the number of atoms in the system
shown in \ref{tab:speedup} suggests that the benefit of MTF increases with the complexity 
of the system size (see Figure \ref{fig:speedup}).
\section{Conclusions}

Mesoscopic systems express behaviors stemming from atom-atom interactions
across many scales in space and time. Earlier approaches based on Langevin
equations for coarse-grained variables did achieve efficiencies over MD
without comprimising accuracy and captured key atomic scale details \cite%
{HierarchicalOP,OPs}. However, such an approach requires the construction of
diffusion factors, a task that consumes significant computational resources.
This is because of the need to use large ensembles and construct correlation
functions.

The multiscale factorization method used here introduces the benefits of
multiscale theory of the LE. Here we revisit the Trotter factorization
method within our earlier multiscale context. A key advantage is that the
approach presented here avoids the need for the resource-consuming diffusion
factors, and thermal average and random forces. The CG variables for the
mesoscopic systems of interest do have a degree of stochastic behavior. In
the present formulation, this stochasticity is accounted for via a series of
MD steps used in the phase of the multiscale factorization algorithm wherein
the $N-$atom probability density is evolved via $\mathcal{L}_{micro}$ , i.e.
at constant value of the CG variables. 

The MTF algorithm can be further optimized to produce greater speedup factors. In particular,
the results obtained here can be significantly improved with the following: 
1) after updating the CGs in the
two-phase coevolution Trotter cycle, one must fine grain i.e. develop the atomistic
configuration to be used as an input to MD. Recently, we have shown that the
CPU time to achieve this fine graining can be dramatically reduced via a
constraint method that eliminates bond length and angle strains,
2) information from
earlier steps in discrete time evolution can be used to increase the
time step and achieve greater numerical stability; while
this was demonstrated for one multiscale algorithm \cite{History}, 
it can also be adapted to the multiscale factorization method, and 3) 
the time stepping algorithm used in this work is the the analogue 
of the Euler method for differential equations, and greater numerical 
stability and efficiency could be achieved for a system of stiff
differential equations using implicit and semi-implicit schemes \cite{Numerical}.

\acknowledgement
This project was supported in part by the National Science Foundation
(Collaborative Research in Chemistry Program), National Institutes of Health
(NIBIB), METAcyt through the Center of Cell and Virus Theory, Indiana
University College of Arts and Sciences, and the Indiana University
information technology services (UITS) for high performance computing resources.

\bibliography{refs}

\begin{figure}[!H]
\centering
\includegraphics[width=\textwidth]{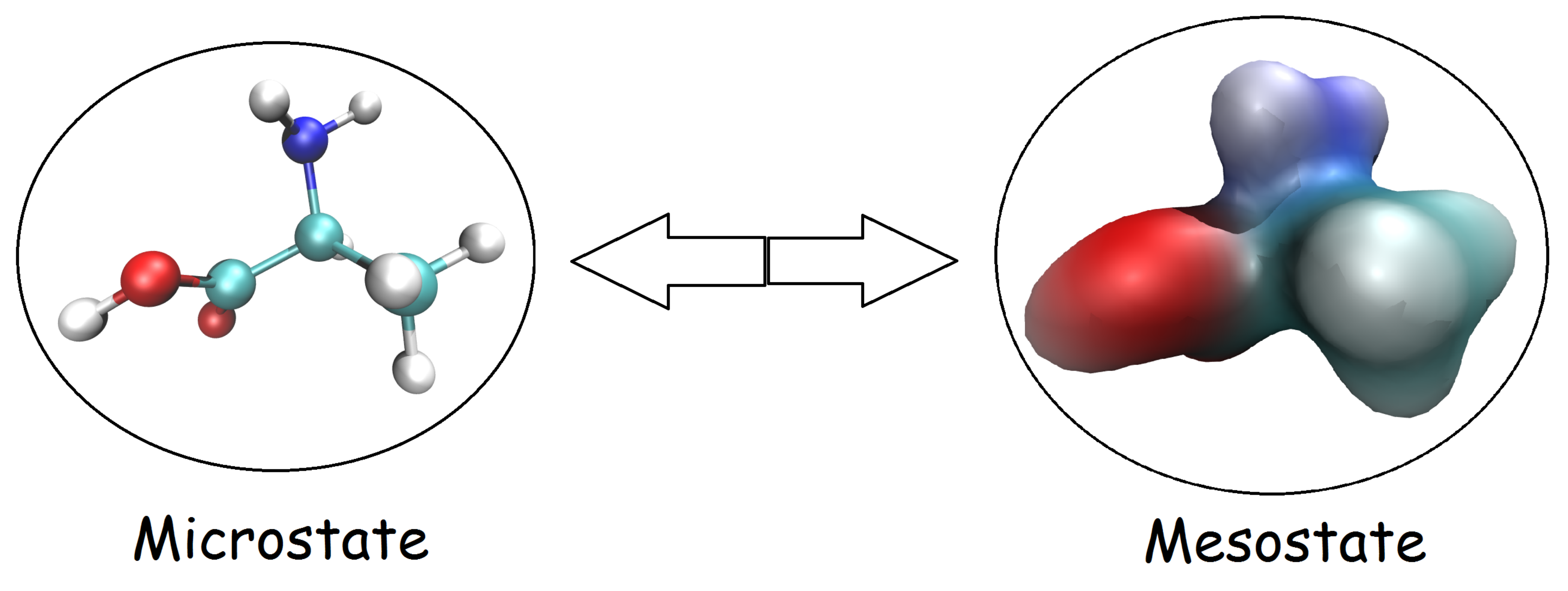}
\caption*{For table of contents use only.}
\end{figure}

\end{document}